# Comment on "Theory of high-force DNA stretching and overstretching"

Pui-Man Lam[*]
Physics Department, Southern University
Baton Rouge, Louisiana 70813

Abstract

Recently Storm and Nelson [1] (Phys.Rev. E67, 51906 (2003)) introduced the discrete persistent chain model which contains both features of the freely jointed chain (FJC) and the wormlike chain (WLC) models. Equation (20) of their paper is correct only in a special case of large $\tilde{l}$, the ratio of the persistence length to the monomer length. This special is unnecessary because the general case can be studied just as easily. Working out the general case, we obtain the force extension relation correct for all values of the parameter $\tilde{l}$. This force extension relation reduces to the FJC result at small $\tilde{l}$ and to the WLC at large $\tilde{l}$. At small force, it reduces to the result of Rosa et al (cond-mat 0307015).

Recently Storm and Nelson [1] proposed a model, the discrete persistent chain (DPC) that borrows features from both the freely jointed chain (FJC) and the wormlike chair (WLC), and show that it resembles the data more closely. Next they elaborate their model to allow coexistence of two confomational states of DNA, each with its own stretch and bend elastic constants. Their elaborated model gives an excellent fit through the entire overstretching transition of nicked, double-stranded DNA. Our comment here concerns only with the first part of their paper, the discrete persistent chain.

Their DPC models the polymer as a chain composed of N segments of length b, whose conformation is fully described by the collection of orientation vectors $\{\hat{t}_i\}$. Bend resistance is taken into account by including an energy penalty at each link proportional to the square of the angle ($\theta_{i,i+1} \equiv \arccos(\hat{t}_i \cdot \hat{t}_{i+1})$) between two subsequent links. In an external force $f\hat{z}$ in the z-direction, the partition function is given by

$$Z = \prod_{i=1}^{N} \int_{S^2} d^2\hat{t}_i\, e^{(fb/2k_BT)\hat{t}_1\cdot\hat{z}} \prod_{i=1}^{N-1} e^{-\beta_i(\hat{t}_i,\hat{t}_{i+1})/k_BT} e^{(fb/2k_BT)\hat{t}_N\cdot\hat{z}} \tag{1}$$

where

$$\frac{\beta_i(\hat{t}_i,\hat{t}_{i+1})}{k_BT} = -\frac{fb}{2k_BT}(\hat{t}_i + \hat{t}_{i+1})\cdot\hat{z} + \frac{A}{2B}(\theta_{i,i+1})^2 \tag{2}$$

and $S^2$ is the two-dimensional unit sphere.

As is shown in ref.[1], for large N, the partition function is given by

$? ? ?_{max}^N$ (3)

where $?_{max}$ is the maximum eigenvalue of the transfer matrix **T** with matrix elements

$?(\hat{t}_i, \hat{t}_j) ? e^{??(\hat{t}_i \cdot \hat{t}_j)/k_B T}$. (4)

Therefore in the limit of large N, the force extension relation is given by

$$\left\langle \frac{z}{L} \right\rangle ? \frac{k_B T}{L} \frac{d}{df} \ln ? ? \frac{k_B T}{L} \frac{d}{df} \ln ?_{max}^N ? \frac{k_B T}{b} \frac{d}{df} \ln ?_{max}$$ (5)

where L is the contour length of the chain.

Using a one parameter trial eigenfunction

$v_?(\hat{t}) ? e^{?\hat{t}\hat{z}}$ (6)

with parameter $?$ and squared norms

$$\|v_?\|^2 ? \frac{2?}{?} \sinh(2?),$$ (7)

the maximum eigenvalue $?_{max}$ can be variationally approximated by

$$?_{max}^* ? \max y(?) ? \max \frac{v_? ?? ?v_?}{\|v_?\|^2}$$ (8)

Evaluating the matrix product, they find that

$$y(?) ? \frac{8?? \cosech(2?)}{\tilde{l}(\tilde{f}?2?)} \exp\!\left[? \frac{3}{2}\tilde{l} ? \frac{(\tilde{f}?2?)^2}{8\tilde{l}}\right] ?\!\int_{\tilde{l}?(\tilde{f}/2??)}^{\tilde{l}?\tilde{f}/2??} dG \exp[G^2/(2\tilde{l})]\sinh(G)$$

(9)

where

$$\tilde{f} ? \frac{fb}{k_B T}, \qquad \tilde{l} ? \frac{A}{b}.$$

This integral can be evaluated in terms of error function. The result is

$$y(\lambda) = \frac{\pi i 2\sqrt{2}\lambda^{3/2}}{\sqrt{\tilde{l}}(2\lambda - \tilde{f})} \exp\left[2\tilde{l}\lambda - \frac{(2\lambda - \tilde{f})^2}{8\tilde{l}}\right] \text{cosech}(2\lambda)$$

$$\times \left\{Erf\left[i\frac{4\tilde{l}\lambda - \tilde{f} - 2\lambda}{2\sqrt{2\tilde{l}}}\right] - Erf\left[i\frac{4\tilde{l}\lambda - (\tilde{f} - 2\lambda)}{2\sqrt{2\tilde{l}}}\right] - 2Erf\left[i\frac{\tilde{f} - 2\lambda}{2\sqrt{2\tilde{l}}}\right]\right\} \quad (10)$$

This result is different from Equation (20) of ref.[1] in which the last error function in (10) is missing. The authors of ref. [1] have stated that their Equation (20) is only valid in the parameter regime where $\lambda^*$ obeys

$$\lambda^* \gg \tilde{l} \gg \frac{\tilde{f}}{2}. \quad (11)$$

This special case is unnecessary since the general expression (10) valid for all parameter values can be studied just as easily..

Expanding (10) to second order in $\lambda$ and f one can analytically calculate $\lambda^*$ to be

$$\lambda^* = \frac{\tilde{f}}{2}\frac{1 - \mathcal{L}(\tilde{l})}{1 + \mathcal{L}(\tilde{l})} \quad (12)$$

where $\mathcal{L}(x) = \coth(x) - x^{-1}$ is the well known Langevin function. Substituting this into the expression for $y(\lambda)$ expanded to second order in $\lambda$ and f, one finds

$$\left\langle \frac{z}{L} \right\rangle = \frac{fb}{3k_B T}\frac{1 - \mathcal{L}(\tilde{l})}{1 + \mathcal{L}(\tilde{l})} \quad (13)$$

This is the force extension relation for the (DPC) valid for small force but for all $\tilde{l}$. It agrees with that of Rosa et al [2] and reduces to the FJC and the WLC results in the limits $\tilde{l} \to 0$ and $\tilde{l} \to \infty$ respectively. It differs from equation (23) of ref.[1] which is valid only for large $\tilde{l}$.

For general f and $\tilde{l}$, analytic expressions for both $dy(\lambda)/d\lambda$ and $d\ln y(\lambda)/df$ can be obtained using (10). For fixed values of f and $\tilde{l}$, the equation $dy(\lambda)/d\lambda = 0$ can be solved numerically for $\lambda^*$. Substituting this value of $\lambda^*$ into the expression for $d\ln y(\lambda)/df$ will give the extension <z/L> for the fixed values of f and $\tilde{l}$. The results obtained are shown in Figure 1, compared with the FJC and the WLC results. This figure shows that even for large force, the DPC reduces to FJC and the WLC results in the limits $\tilde{l} \to 0$ and $\tilde{l} \to \infty$ respectively. This shows that the DPC model interpolates between the FJC and the WLC with the parameter $\tilde{l}$.

* email: pmlam@grant.phys.subr.edu

Figure Caption:
Figure 1: Average extension <z/L> versus reduced force $fb/(k_BT)$, for various models. The symbols represent numerical solution of the discrete persistent chain (DPC) for various values of the parameter $\tilde{l}$. The solid line denotes the freely jointed chain model (FJC). The broken lines denote the worm like chain (WLC) model for various values of the parameter $\tilde{l}$.

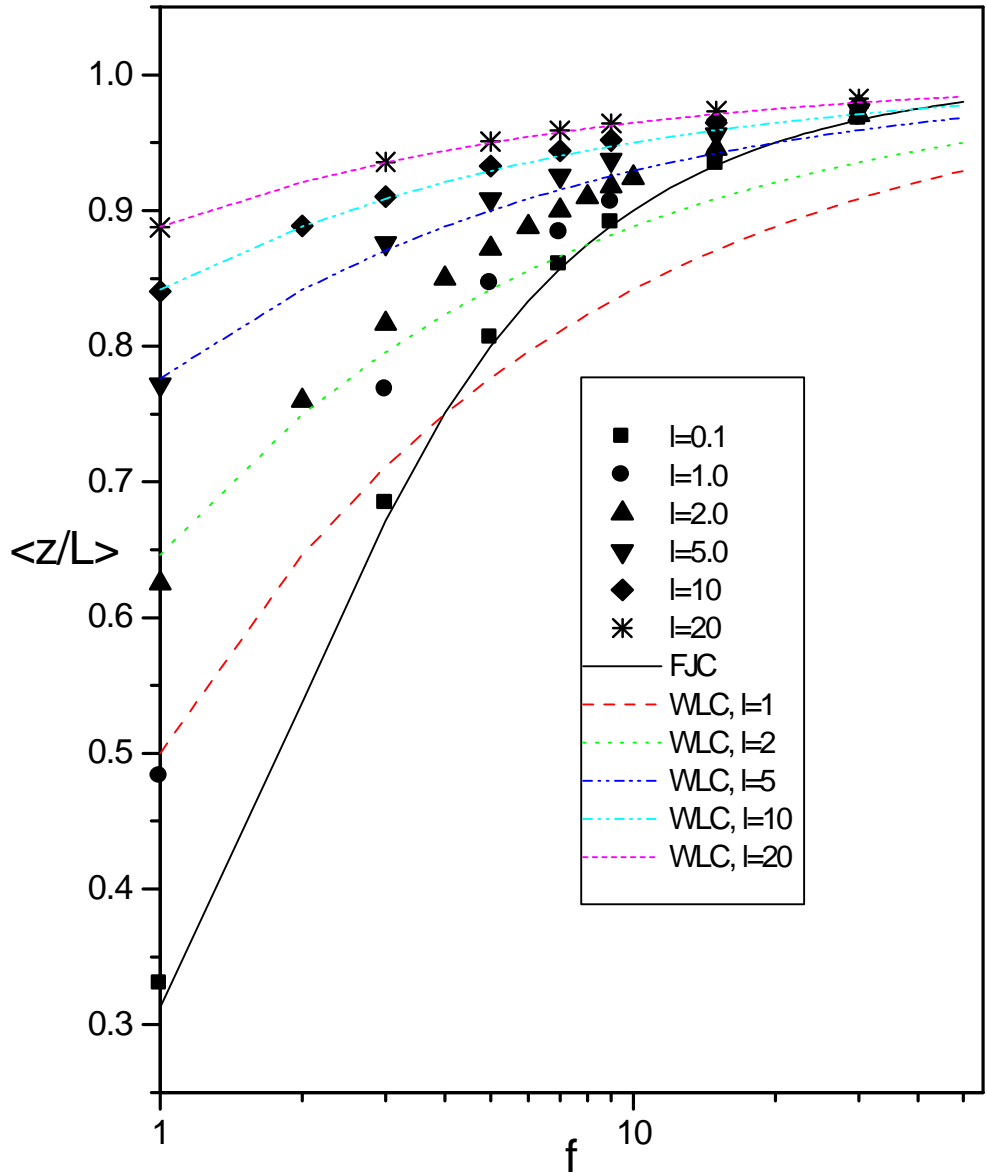